\documentstyle[amssymb,prl,aps,twocolumn,epsf]{revtex}
\begin{document}
\draft
\title{Localization-delocalization transition in disordered systems with a direction}
\author{A.~V.~Kolesnikov$^{1}$ and K.~B.~Efetov$^{1,2}$}
\address{$^{1}$Fakult\"{a}t f\"{u}r Physik und Astronomie, Ruhr-Universit\"{a}t\\
Bochum, Universit\"{a}tsstr. 150, Bochum, Germany\\
$^{2}$L.D. Landau Institute for Theoretical Physics, Moscow, Russia}
\date{\today{}}
\maketitle

\begin{abstract}
Using the supersymmetry technique, we study the loca\-liza\-tion-delo\-cali\-za\-tion
transition in quasi-one-dimensional non-Hermitian systems with a direction.
In contrast to chains, our model captures the diffusive character of
carriers' motion at short distances. We calculate the joint probability of
complex eigenvalues and some other correlation functions. We find that the
transition is abrupt and occurs as a result of an interplay between two
saddle-points in the free energy functional. 
\end{abstract}


%\begin{multicols}{2}

Non-Hermitian models with disorder have attracted recently a considerable
attention \cite{Nelson,weak,EfnoH,chain,lerner,Zee,all,Yurk,HyDy,QCD,QMech}. 
%One comes to non-Hermitian Hamiltonians reducing a $d+1$ dimensional classical 
%system to a $d$-dimensional quantum one \cite{Nelson}. They also 
Non-Hermitian Hamiltonians appear in context of flux lines in superconductors
\cite{Nelson}, in transfer phenomena in lossy media \cite{Yurk}, in hydrodynamics
\cite{HyDy}, in QCD \cite{QCD}, in quantum mechanics of open systems \cite{QMech}.

The most important property of the non-Hermitian Hamiltonians is that their
eigenenergies can be complex independently of their origin.
%type of the non-Hermiticity.
Very interesting physical systems are models with a direction. The simplest 
Hamiltonian ${\cal H}$ of such models is written as 
\begin{equation}
{\cal H}=\frac{({\bf p}+i{\bf h})^{2}}{2m}+V({\bf r}),  \label{a1}
\end{equation}
where ${\bf p}$ is the momentum operator, $V\left( {\bf r}\right) $ is a
random potential and ${\bf h}$ is a constant vector.
One comes to the Hamiltonian ${\cal H}$, Eq.~(\ref{a1}), reducing the $d+1$
dimensional classical problem of vortices in a superconductor with columnar
defects to a $d$-dimensional quantum problem. Then, the vector ${\bf h}$ is
proportional to the component of the magnetic field perpendicular to the
line defects \cite{Nelson}. 
%One can also imagine other physical systems for
%which the Hamiltonian ${\cal H}$ could be relevant.

Considering the one-dimensional (1D) version of the Hamiltonian ${\cal H}$, Eq.~(\ref{a1}),
the authors of Ref. \cite{Nelson} predicted a localization-delocalization
transition that occurs at $h_{c}=l_{c}^{-1}$, where $l_{c}$ is
the localization length at ${\bf h=}0$. At $h\leq l_{c}^{-1}$
the ``imaginary vector-potential'' ${\bf h}$ can be removed by the ``gauge
transformation'' 
\begin{equation}
\phi _{k}\left( {\bf r}\right) =\phi _{0k}\left( {\bf r}\right) \exp \left( 
{\bf hr}\right) ,\;\bar{\phi}_{k}\left( {\bf r}\right) =\phi _{0k}\left( 
{\bf r}\right) \exp \left( -{\bf hr}\right)  \label{a0}
\end{equation}
In Eq.~(\ref{a0}), $\phi _{k}\left( {\bf r}\right) $ and $\bar{\phi}%
_{k}\left( {\bf r}\right) $ are the right and left eigenfunctions, $\phi
_{0k}\left( {\bf r}\right) $ are those at ${\bf h=}0$. Carrying out
the transformation, Eq.~(\ref{a0}), we see that the eigenvalues of the
localized states $\epsilon _{k}$ do not depend on ${\bf h}$.
%coincide with the eigenvalues of the Hamiltonian at ${\bf h=}0$. 
However, the gauge transformation applies
only if the functions $\phi _{k}\left( {\bf r}\right) $ do not grow at
infinity. For eigenfunctions of the form $\phi _{0}\sim \exp (-|{\bf r|}%
/l_{c})$, this is possible if $h<h_{c}$. For $h>h_{c}$, one should take
extended wave functions, which leads to complex eigenenergies.

Qualitative arguments of Ref. \cite{Nelson} have been confirmed for the
disordered chains numerically as well as analytically \cite{chain,lerner,Zee}%
. At the same time, they are quite general and one might expect that such a
%the localization-delocalization 
transition occurs in more complex systems %with a direction 
provided all states at ${\bf h=}0$ are localized. For example, one can think of 
disordered quasi-one-dimensional wires and films. Using the
mapping of Ref. \cite{Nelson}, one expects such models to be relevant
for description of the vortices in slabs and 3D bulk superconductors.
%Theoretically, t
The wires and films are richer than the chains because at
distances larger than the mean free path $l$ but smaller than the
localization length $L_{c}$ the motion is diffusive, whereas in disordered
chains only ballistic motion or localization are possible. As a result,
properties of the localization-delocalization transition in wires and films
can differ. The problem has not been addressed yet and our paper is aimed to 
provide its basic understanding. % of properties of the transition.

%Below, w
We study the distribution function of complex eigenenergies $P\left(
\varepsilon ,y\right) $, where $\varepsilon $ and $y$ are the real and
imaginary parts of the energy, and correlations of eigenfunctions of the
Hamiltonian ${\cal H}$, Eq.~(\ref{a1}), at different points for wires using
the supersymmetry technique \cite{book}. A proper non-linear supermatrix $%
\sigma $-model has been derived in Ref. \cite{EfnoH}.
%and one can try to study this model by the transfer matrix technique. The 
Its free energy functional $F_{h}\left[ Q\right] $ 
%of the non-linear $\sigma $-model  is written as 
reads \begin{equation}
F_{h}\left[ Q\right] =\frac{\pi \nu }{8}STr\int [D\left( \partial Q\right)
^{2}-4\left( \gamma \Lambda +y\Lambda _{1}\tau _{3}\right) Q]d{\bf r}
\label{a6}
\end{equation}
where $D$ is the classical diffusion coefficient, $\nu $ is the density of
states, and the standard notations for the supertrace $Str$ and
supermatrices $Q$, $\tau _{3}$, $\Lambda $, and $\Lambda _{1}$ are used \cite
{book,EfnoH}. The free energy functional $F_{h}\left[ Q\right] $,  Eq.~(\ref
{a6}), differs from the conventional one $F_{0}\left[ Q\right] $ (describing
the Hermitian problem) by the presence of the ``gauge-invariant''
combination ${\bf \partial }Q=$ $\nabla Q+{\bf h}\left[ Q,\Lambda \right] $
instead of $\nabla Q$.

The zero-dimensional (0D) version of the $\sigma $-model is equivalent to
models of random non-Hermitian matrices %and has been studied in Ref. 
\cite{EfnoH}, where no transition occurs. But the $\sigma $-model, Eq. 
(\ref{a6}), holds in any dimension and one might expect that already
the 1D version would manifest the %localization-delocalization 
transition. So, one could describe the transition using the transfer-matrix 
technique developed for the $\sigma $-models \cite{EfLa,Kol}. 

%It comes as a great surprise that 
Surprisingly, the 1D $\sigma $-model with $F_{h}\left[ Q%
\right] $,  Eq.~(\ref{a6}), {\it does not} have any transition and physical
quantities of interest are smooth functions of $h$ giving always a finite
probability of complex eigenenergies. At first glance, this result is in an
evident contradiction with the arguments of Ref. \cite{Nelson}. The
resolution of this paradox is quite interesting. It turns out that the $%
\sigma $-model with $F_{h}[Q]$,  Eq.~(\ref{a6}), describes properly only the
region of the delocalized states. 
%As concerns the localized states, one should use t
The functional $F_{0}[Q]$ should be used in the localized regime, for any 
finite $h<h_{c}$, where $h_{c}$ is a critical vector-potential. This 
replacement of the free energy, resembling a first order transitions, 
%in the theory of phase transitions, 
results in an {\it abrupt} change of the distribution of eigenvalues. 
At $h<h_{c}$, all eigenenergies are real whereas at $h>h_{c}$ one gets a 
broad distribution in the complex plane. In the limit 
$h\gg h_{c}$, our
results are universal for any dimension. All this is in a contrast with the
results obtained for the chains, where %the analysis of the spectrum leads to
%the conclusion that the transition occurs
the spectrum changes smoothly, showing the appearance of
``arcs'' when increasing $h$ \cite{Nelson,chain}, or ``wings'' when starting
from the regime of the strong non-Hermiticity \cite{Zee}.

Having formulated the rough picture of what happens when changing the
parameter $h$, let us present some details of calculations. We first
introduce the quantities which are to be studied. Since eigenvalues of the
Hamiltonian (\ref{a1}) can be complex, it is convenient to double the size
of relevant matrices \cite{EfnoH}, thus ``hermitizing'' the problem. In such
an approach, the role of the Green function is played by the function 
\begin{equation}
B({\bf r},{\bf r}^{\prime })=\sum_{k}\frac{\gamma \phi _{k}({\bf r})%
\overline{\phi }_{k}({\bf r}^{\prime })}{(\epsilon -\epsilon _{k}^{\prime
})^{2}+(y-\epsilon _{k}^{\prime \prime })^{2}+\gamma ^{2}}\;  \label{a2}
\end{equation}
with an eigenenergy $\epsilon _{k}=\epsilon _{k}^{\prime }+i\epsilon
_{k}^{\prime \prime }$. The ``density-density correlator'' in the present
context is given by 
\begin{equation}
Y({\bf r},{\bf r}^{\prime };\epsilon ,y)=\frac{1}{\pi }\lim_{\gamma
\rightarrow 0}\,\langle B({\bf r},{\bf r}^{\prime })B({\bf r}^{\prime },{\bf %
r})\rangle \;,  \label{a3}
\end{equation}
where brackets imply impurity averaging. % and $V$ is the volume. 
The limit $\gamma \rightarrow 0$, understood in all correlators, becomes 
important as soon as $h\rightarrow 0$. The function $Y({\bf r},{\bf r}^{\prime 
};\epsilon,y)$,  Eq.~(\ref{a3}), establishes a link between the localization 
properties and the joint probability density of complex eigenenergies 
\begin{equation}
P(\epsilon ,y)=\frac{1}{V}\int {\rm d}{\bf r}\,{\rm d}{\bf r}^{\prime }\,Y(%
{\bf r},{\bf r}^{\prime };\epsilon ,y)\;  \label{a4}
\end{equation}
where $V$ is the volume.
We introduce also another important correlator $X\left( {\bf r,r}^{\prime
};\epsilon ,y\right) =C\left( {\bf r,r}^{\prime };\epsilon ,y\right)
+C\left( {\bf r}^{\prime },{\bf r;}\epsilon {\bf ,}y\right) ,$ 
\begin{equation}
C\left( {\bf r,r}^{\prime };\epsilon ,y\right) =\frac{1}{2\pi }\lim_{\gamma
\rightarrow 0}\,\langle B({\bf r},{\bf r}^{\prime })B^{\ast }({\bf r},{\bf r}%
^{\prime })\rangle  \label{a5}
\end{equation}
The correlator $Y({\bf r},{\bf r}^{\prime };\epsilon ,y)$,  Eq.~(\ref{a3}),
is invariant under the transformation,  Eq.~(\ref{a0}), but $X\left( {\bf r,r}%
^{\prime };\epsilon ,y\right) $ is not.

We further express as usual \cite{EfnoH,book} the correlation functions in 
terms of integrals over eight-component superfield $\psi ({\bf r})$,
average over the white-noise disorder potential $V({\bf r})$, decouple $\psi
^{4}$ term by $8\times 8$ matrix $Q({\bf r})$ and integrate over $\psi
\left( {\bf r}\right) $. As a result, we obtain a functional integral over $%
Q $ with a free energy functional %$F[Q]$ that can be written as 

\[
F[Q]=\frac{1}{2}\int STr\left( \frac{\pi \nu Q^{2}({\bf r)}}{4\tau }-\ln [-i{\cal %
H}+\frac{Q({\bf r}) }{2\tau }] \right)d{\bf r} 
\]
\begin{equation}
%{\cal H}_{00}=H_{0}^{\prime }-\epsilon +i\gamma \Lambda ,\quad {\cal H}%
%_{01}=i\Lambda _{1}(H^{\prime \prime }+y\tau _{3})  \label{a50}
%\end{equation}
%where $H_{0}^{\prime }=\frac{{\bf p}^{2}-{\bf h}^{2}}{2m}$, $H^{\prime
%\prime }=-i\frac{{\bf h\nabla }}{m}$ and $\tau $ is the mean free time.
{\cal H}=({\bf p}^{2}-{\bf h}^{2})/2m-\epsilon +i\gamma \Lambda + i\Lambda _{1}
{\cal H}^\prime 
\label{a50} \end{equation}
where ${\cal H}^\prime=-i {\bf h\nabla}/m+y\tau_3$ and $\tau $ is the mean free time.

The next standard step is to find the minimum of $F[Q]$ neglecting ${\cal H}^\prime$. 
The minimum is reached at  
\begin{equation}
Q=V\Lambda \overline{V} \;  \label{a500} \end{equation} 
$V\overline{V}=1$, $V^{+}=K\overline{V}K$. (The notations are the same as in Ref. 
\cite{EfnoH,book}). Expanding the functional $F[Q]$ near the minimum in the
gradients of $Q$ and in ${\cal H}^\prime$ one comes to the functional $F_{h}[Q]$,
 Eq.~(\ref{a6}). This is exactly the way how the $\sigma $-model,  Eq.~(\ref
{a6}), was derived in Ref. \cite{EfnoH}. However, it has been mentioned that
this $\sigma $-model is not valid in the localized regime. What is
wrong in the derivation?

It not difficult to find that the functional $F\left[ Q\right] $ has also
another minimum. Performing the transformation
\begin{equation}
\widetilde{Q}=\exp \left( \Lambda _{1}{\bf rh}\right) Q\exp \left( -\Lambda _{1}%
{\bf rh}\right)  \label{a51}
\end{equation}
we rewrite the functional $F\left[ Q\right] ,$  Eq.~(\ref{a50}) in terms of $%
\widetilde{Q}$. As a result, the imaginary vector-potential ${\bf h}$ is removed
from $F$ and the minimum is achieved at 
\begin{equation}
\widetilde{Q}=V\Lambda \overline{V}  \label{a52}
\end{equation}
which corresponds to $Q$ varying in space.

Which of these two saddle points should be chosen? The answer depends on the
value of $h$. To clarify this question we consider the correlation functions 
$Y\left( {\bf r,r}^{\prime };\epsilon ,y\right) $ and $X\left( {\bf r,r}%
^{\prime };\epsilon ,y\right) $, Eqs. (\ref{a3},\ref{a5}). They can be
written as functional integrals over the supermatrices $Q$
\begin{equation}
Y\left( X\right) =\frac{\pi \nu ^{2}}{4}\,\,\left\langle Q_{42}^{1\pm
}Q_{24}^{\prime 1\pm }-Q_{42}^{2\pm }Q_{24}^{\prime 2\pm }\right\rangle _{Q}
\label{a7}
\end{equation}
where the sign $+$ ($-$) corresponds to the correlator $Y$ $\left( X\right) $%
, $Q^{1\pm }=Q^{11}({\bf r})\pm Q^{22}({\bf r})$, $Q^{2\pm }=Q^{12}({\bf r}%
)\pm Q^{21}({\bf r})$, and the matrices $Q^{\prime }$ are taken at ${\bf r}%
^{\prime }$. The symbol $\left\langle ...\right\rangle _{Q}$ stands for
averaging with the functional $F\left[ Q\right] $,  Eq.~(\ref{a50}). To
calculate the integral in  Eq.~(\ref{a7}) let us use the transformation, Eq.
(\ref{a51}), and take the saddle-point,  Eq.~(\ref{a52}). Since the 
combination of the supermatrices $Q$ entering  Eq.~(\ref{a7}) for the 
function $Y$ % \left( {\bf r,r}^{\prime };\epsilon ,y\right) $ 
is invariant under the transformation, Eq.~(\ref{a51}), ${\bf h}$ is gauged
out in this function. Therefore, one can use the standard results of the 
transfer-matrix approach \cite{book} developed for the Hermitian case. 
The final result for the correlator $Y\left( {\bf r,r}^{\prime };
\epsilon ,y\right)$ 
\begin{equation}
Y\left( {\bf r,r}^{\prime };\epsilon ,y\right) =p_{\infty }\left(
r\right) \delta \left( y\right)  \label{a70}
\end{equation}
where $r=\left| {\bf r-r}^{\prime }\right|$, contains the function $p_{\infty
}\left( r\right) $ characterizing localization properties
\begin{equation}
p_{\infty }\left( r\right) =\sum_{k}\left| \phi _{0k}\left( {\bf r}\right)
\right| ^{2}\left| \phi _{0k}\left( {\bf r}^{\prime }\right) \right|
^{2}\delta \left( \epsilon -\epsilon _{k}\right)  \label{a71}
\end{equation}
In disordered wires, the function $p_{\infty }\left( r\right)$ 
describes the decay of the wave functions at $r\gg L_{c}$ 
\begin{equation}
p_{\infty }\left( r\right) \approx \frac{1}{4\sqrt{\pi }L_{c}}\left( \frac{%
\pi ^{2}}{8}\right) ^{2}\left( \frac{4L_{c}}{r}\right) ^{3/2}\exp \left( -%
\frac{r}{4L_{c}}\right)  \label{a72}
\end{equation}
where $L_{c}$ is the localization length. For the unitary ensemble it equals 
$L_{c}=2\pi \nu SD$, $S$ is the cross-section.

In contrast, the vector-potential ${\bf h}$ enters explicitly the pre-exponential
of the function $X$ after making the transformation,  Eq.~(\ref{a51}). 
%the combination for the function $X\left( {\bf r,r}^{\prime };\epsilon ,y\right) 
%$ in  Eq.~(\ref{a7}) contains the vector-potential ${\bf h}$. 
However, the dependence on ${\bf h}$ is simple and calculations for $X\left( 
{\bf r,r}^{\prime };\epsilon ,y\right) $ can be performed in the same way as 
for $Y\left( {\bf r,r}^{\prime };\epsilon ,y\right)$ yielding 
\begin{equation}
X\left( {\bf r,r}^{\prime };\epsilon ,y\right) =\cosh \left( 2hr\right)
p_{\infty }\left( r\right) \delta \left( y\right)  \label{a73}
\end{equation}
Eqs.~(\ref{a70},\ref{a73}) demonstrate that, provided one may perform the
transformation,  Eq.~(\ref{a51}), all eigenenergies remain real. However, the
validity of this procedure depends on the value of $h$. 
The function $X\left( {\bf r,r}^{\prime};\epsilon ,y\right) $ does 
not grow at infinity only if $h<h_{c}$ where
\begin{equation}
h_{c}=\left( 8L_{c}\right) ^{-1}  \label{a74}
\end{equation}
At $h>h_{c}$ Eqs.~(\ref{a70},\ref{a73}) cannot be used because this would
correspond to growing wave functions, which are forbidden for a
closed geometry. This agrees with the arguments of Ref. \cite{Nelson}.
According to Ref. \cite{Nelson} one should use at $h>h_{c}$ extended states
having complex eigenenergies.

In the present formalism, the other saddle-point, Eq.~(\ref{a500}),
of the free energy $F[Q]$ should be taken in the regime $h>h_c$.
%one should take the saddle-point in the functional 
%$F\left[ Q\right] $,  Eq.~(\ref{a50}), given by  Eq.~(\ref{a500}). 
This leads to the $\sigma $-model in the form of  Eq.~(\ref{a6}). 
Expecting that the eigenenergies become complex, 
we determine their distribution function $P\left( \epsilon ,y\right) $, Eq.~(\ref{a4}).
Following the transfer-matrix technique \cite{EfLa,book} we write the function $P\left(
\epsilon ,y\right) $ in the form
\begin{equation}
P\left( \epsilon ,y\right) =\frac{\pi \nu ^{2}S}{4}\int \Psi \left( Q\right)
\,(Q_{42}^{1+}P_{24}^{1+}-Q_{42}^{2+}P_{24}^{2+})\,{\rm d}Q\,  \label{a8}
\end{equation}
%where $S$ is the cross-section of the wire. 
In  Eq.~(\ref{a8}), $\Psi \left(Q\right) $ is the partition function of the 
semi-infinite wire, the matrix function $P$ is the partition function between the 
points $r$ and $r^{\prime}$ multiplied by $Q^{\prime}\Psi(Q^{\prime})$ and 
integrated over $r^{\prime }$. As usual, proper differential equations for $\Psi$ 
and $P$ are derived from $F[Q]$, Eq.~(\ref{a6}).

In order to carry out these calculations it is necessary to choose a
parametrization of the supermatrices $Q$. For the non-Hermitian problem
involved the parametrization of Ref.~\cite{EfnoH} is most natural. For
simplicity we consider the unitary ensemble, where the supermatrices $Q$
can be parametrized in the form 
\[
Q=T\left( \hat{\theta}\right) Q_{0}\bar{T}\left( \hat{\theta}\right) \text{, 
}Q_{0}=\left( 
\begin{array}{cc}
\cos \hat{\varphi} & -\tau _{3}\sin \hat{\varphi} \\ 
-\tau _{3}\sin \hat{\varphi} & -\cos \hat{\varphi}
\end{array}
\right) , 
\]
\begin{equation}
\hat{\varphi}=\left( 
\begin{array}{cc}
\varphi & 0 \\ 
0 & i\chi
\end{array}
\right) \text{, \ }\hat{\theta}=\left( 
\begin{array}{cc}
\theta & 0 \\ 
0 & i\theta _{1}
\end{array}
\right)  \label{a81}
\end{equation}
In Eqs.~(\ref{a81}), the supermatrices $T( \hat{\theta}) $
contain not only real variables $\hat{\theta}$ but also Grassmann variables.
The ``angles'' %$\hat{\varphi}$ and $\hat{\chi}$ 
vary in the following intervals %\begin{eqnarray}
$-\pi /2<\varphi <\pi /2$,
$-\infty <\chi <\infty$,  %\text{, }
$-\pi <\theta <\pi$,
$-\infty <\theta _{1}<\infty$.
%\text{,}\label{a82} \\ \text{ }  \text{, \ }
%\nonumber \end{eqnarray}

The variables $\hat{\theta}$ and $\hat{\varphi}$ are not equivalent. For
example, neglecting the gradient terms in  Eq.~(\ref{a6}) one comes to the 0D
free energy $F^{\left( 0\right) }$, containing the variables $\hat{\varphi}$
only 
\begin{equation}
F^{\left( 0\right) }[\hat{\varphi}]=\tilde{h}^{2}(\lambda _{1}-i\tilde{y}/2%
\tilde{h}^{2})^{2}+\tilde{h}^{2}(\lambda +\tilde{y}/2\tilde{h}^{2})^{2}\,,
\label{a9}
\end{equation}
where $\lambda _{1}=\sinh \chi $, $\lambda =\sin \varphi $, $\tilde{h}%
^{2}=2h^{2}L_{c}^{2}$ and $\tilde{y}=2yL_{c}^{2}/D$. In contrast to real
magnetic field $H$, gradually suppressing and finally freezing out some
degrees of freedom with increasing $H$, the imaginary magnetic field $h$
shifts the saddle point as a whole. Noticeable changes in behavior occur
only at $\pm \tilde{y}\gtrsim 2\tilde{h}^{2}$, where $\sin\varphi\approx 1$.

Calculations performed in Ref.~\cite{EfnoH} for the 0D case show that the
variables $\hat{\theta}$ play a minor part in 0D. They do not enter $%
F^{\left( 0\right) }[\hat{\varphi}]$ but their role is even less pronounced
due to the singularity of the Jacobian at $\theta $, $\theta _{1}\rightarrow
0$. A detailed discussion of Ref.~\cite{EfnoH} leads to the conclusion
that one should replace the $\hat{\theta}$-dependent part of the Jacobian by
a constant and put everywhere else $\hat{\theta}=0$. It is also interesting
to note that the free energy $F_{h}\left[ Q\right] $,  Eq.~(\ref{a6}), is not
invariant against the replacement,  Eq.~(\ref{a51}). Using the
parametrization, Eqs.~(\ref{a81}), one obtains that this replacement
leads to the shift $\widetilde{\theta }=\theta -2i{\bf rh}$, $\widetilde{%
\theta _{1}}=\theta _{1}-2{\bf rh}$. This shift changes the contour of the
integration over $-\pi<\theta<\pi $ %, Eqs. (\ref{a82}), 
in a complicated way, thus demonstrating the violation of the ``gauge symmetry''.

Effective disappearance of the variables $\hat{\theta}$ shows that the correspondence between 
the problem under consideration and random flux models \cite{Alt} is rather obscure because
in the latter, the variables $\hat{\theta}$ play a crucial role. For the quantities we have
studied, Eqs.~(\ref{a3},\ref{a4},\ref{a5}), this mapping is inadequate. 

As in the 0D case, one should put everywhere $\theta $, $\theta _{1}=0$ when
deriving the transfer matrix equations. As a result, 
differential equations for $\Psi $ and $P$ contain $\lambda $ and $%
\lambda _{1}$ only 
\begin{eqnarray}
\widehat{{\cal H}}\Psi &=&F^{\left( 0\right) }\Psi ,\qquad \,\widehat{{\cal H%
}}P^{+}=L_{c}(i\lambda _{1}-i\lambda )\Psi ,\,  \label{a10} \\
\widehat{{\cal H}} &=&\frac{1}{J_{0}}\partial _{\lambda }(1-\lambda
^{2})J_{0}\partial _{\lambda }+\frac{1}{J_{0}}\partial _{\lambda
_{1}}(1+\lambda _{1}^{2})J_{0}\partial _{\lambda _{1}}\,\,,  \nonumber
\end{eqnarray}
where $J_{0}=1/(\lambda _{1}+i\lambda )^{2}$.

Analysis of Eqs.~(\ref{a10}) for $hL_{c}\gg 1$ is to some extent similar to
the one in the limit of high frequencies for the conventional problem of localization 
\cite{book}. We obtain 
\[
\Psi \simeq \exp [-\tilde{h}(\lambda _{1}+i\tilde{y}/2\tilde{h}^{2})^{2}-%
\tilde{h}(\lambda -\tilde{y}/2\tilde{h}^{2})^{2}],\;P^{+}\simeq \Psi /2 \tilde{h}
\]
which gives after substitution into  Eq.~(\ref{a8}) 
\begin{equation}
P(\epsilon ,y)\simeq \frac{\nu }{4\tilde{h}^{2}}\left\{ 
\begin{array}{ccr}
1\,, &  & |\tilde{y}|<2\tilde{h}^{2} \\ 
0\,, &  & |\tilde{y}|>2\tilde{h}^{2}
\end{array}
\right. \,  \label{a11}
\end{equation}
The form of the function $P(\epsilon ,y)$,  Eq.~(\ref{a11}), is the same as
the 0D result of Refs. \cite{weak,EfnoH}. This result does not depend on
the dimensionality and corresponds to the elliptic law for strongly
non-Hermitian random matrices \cite{strong}.

Analogous calculations for $Y$,  Eqs.~(\ref{a2}, \ref{a3}) yield 
\begin{equation}
Y(r)\simeq \nu \beta \exp (-2\beta |r|)\,,\,\beta =\sqrt{%
(h^{2}-y^{2}/4Dh^{2})}\,.  \label{a12}
\end{equation}

Analytical study of Eqs.~(\ref{a10}) is hardly feasible at $hL_{c}\sim 1$.
To solve them numerically, we use the standard over-relaxation
method with Chebyshev acceleration. The results for the distribution $%
P(\epsilon ,y)$ are presented in Fig.~1 for several values of $\tilde{h}$.
The lowest value of $\tilde{h}=1/4\sqrt{2}$  corresponds to the
critical $h_{c}$,  Eq.~(\ref{a74}). For comparison we present the 0D
result of Refs. \cite{weak,EfnoH}. 

The curves for the 1D and 0D cases are rather close to each other. However,
their evolution with decreasing $h$ is drastically different. The 0D curve
tends smoothly to the $\delta $-function $\nu \delta \left( y\right) $
when $h\rightarrow 0$,  wheareas\ the 1D curve changes abruptly to this 
expression at $h=h_{c}$. In the region $h>h_{c}$ the states are extended and 
the exponential decay of the correlation function $Y\left( r\right) $,  
Eq.~(\ref {a12}), is obtained after summation over many states with different 
phase differences. % between the points $r$ and $r^{\prime }$.  

\begin{figure} \narrowtext {
\caption{The joint distribution function of complex eigenvalues $P(\epsilon)$,
Eqs.~(\ref{a4}), for the quasi-1D infinite wire (solid line) and the 0D case
(dashed line), for $\tilde{h}=\tilde{h}_c$,0.2,0.25.}
\epsfysize =10.1cm
\centerline{\epsfbox{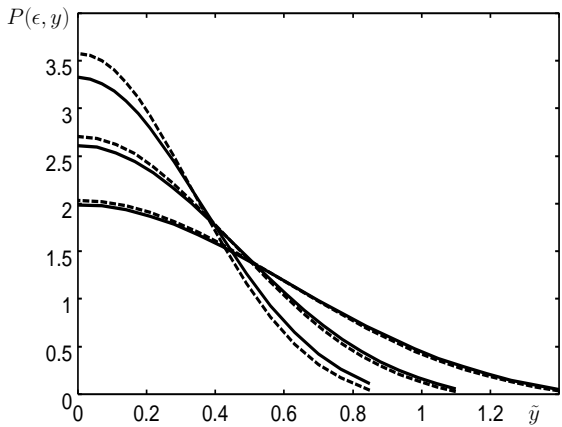}}}
\label{fig1}  
\end{figure}

\vspace*{-4cm}
In conclusion, we studied the loca\-liza\-tion-delo\-ca\-liza\-tion transition for
non-Hermitian disordered systems with a direction,  Eq.~(\ref{a1}) and broken
time-reversal symmetry. We found that the transition occurs due to an
interplay between two saddle points in the free energy $F[Q]$, which results
in the {\it abrupt} transition at the critical field $h=h_{c}.$ Below this
field, all eigenvalues are real and wave functions are localized. Above the
critical field the eigenvalues form a broad distribution in the complex
plane. We believe that the abrupt transition found is quite natural for the
original problem of the vortex lines in the presence of columnar defects 
\cite{Nelson}. Provided the geometry of the sample is closed and it is
infinitely long, vortex lines should abruptly change their behavior, from
being pinned by columns to become oriented along the magnetic field.

We acknowledge useful discussions with I.L.~Aleiner, A.~Altland, C.W.J.~Beenakker
and Y.V.~Fyodorov. This work was supported by SFB 237 ``{\it Unordnung und Grosse
Fluktuationen''}

%\end{multicols}

\end{document}